# Influence of Publication Capacity on Journal Impact Factor for International Open Access Journals from China: Insights from Microeconomic Analysis

Xinyi Chen*

*Editorial Office of Electrochemical Energy Reviews, Periodicals Agency of Shanghai University, Shanghai University, Shanghai 20444, China*

*Coresponding author's Email address: shirleychen@i.shu.edu.cn
Orcid of Xinyi Chen: 0000-0003-2654-4308

**Abstract**
The evolving landscape of open access (OA) journal publishing holds significant importance for policymakers and stakeholders who seek to make informed decisions and develop strategies that foster sustainable growth and advancements in open access initiatives within China. This study addressed the shortcomings of the current journal evaluation system and recognized the necessity of researching the elasticity of annual publication capacity (PUB) in relation to the Journal Impact Factor (JIF). By constructing an economic model of elasticity, a comparative analysis of the characteristics and dynamics of international OA journals from China and overseas was conducted. The analysis categorized OA journals based on their respective elasticity values and provided specific recommendations tailored to each category. These recommendations offer valuable insights into the development and growth potential of both OA journals from China and overseas. Moreover, the findings underscore the importance of strategic decision-making to strike a balance between quantity and quality in OA journal management. By comprehending the dynamic nature of elasticity, China can enhance its OA journal landscape, effectively meet the academic demand from domestic researchers, minimize the outflow of OA publications to overseas markets, and fortify its position within the global scholarly community.

## 1. Introduction

Currently, China encourages the establishment of open access (OA) English-language academic journals and maintains relatively lenient regulations. It provides substantial funding and policy support for international cooperation and OA publishing. In 2022, there was a 44% increase in the number of OA articles published by the Chinese authors as corresponding authors, and a 31% increase in the number of OA publications funded by the National Natural Science Foundation of China compared to the previous year. However, the Chinese publishers have heavily relied on international cooperation in journal publishing, which raises concerns about intellectual property rights and academic freedom(Chen, 2023), ultimately impacting their journal publishing capabilities and academic influence. To address this, China is formulating an Open Science Action Plan. Therefore, it is crucial to conduct research on the development of China's international OA journals in order to assess their current status, identify areas for improvement, and determine future directions.

The report of "Open Access Publishing in China (2022)"(CAST & STM, 2022) jointly compiled by the China Association for Science and Technology (CAST) and the International Association of Scientific, Technical, and Medical Publishers (STM), is comprehensive in describing all aspects of China's open access practices; however, its statistical part mainly is from the perspective of the Chinese researchers, and focuses on international OA publications from China and the constitution of China's OA journals, without considering the publishing capabilities of China's OA journals in terms of meeting the practical needs of the Chinese researchers. Previous studies conducted by Ren et al.(Bi Ning, Du, REN, & Chen, 2021; REN, LI, YANG, Ning, & Chen, 2022; REN, Ning, Chen, & Chen, 2020; REN, Ning, & Yan, 2018; REN et al., 2019; REN, YANG, Ning, Chen, & Ma, 2023) conducted statistical analysis of the development of English-language scientific journals in China. Qian (QIAN, 2022) analyzed the global and China's publication trends in the Web of Science (WoS) Core Collection, including the number of published papers and OA papers. However, there was a lack of in-depth analysis regarding the gap between the Chinese researchers' OA publishing demand and the annual capacity of OA publications in China's international OA journals.

Among the few scientometric studies on China's OA journals, the research focus is more directed at one aspect or one research field. Ding et al. (DING & LI, 2022) primarily focused on the journal impact factor (JIF), quartile rankings, and article processing charges (APCs) of OA journals, suggesting that China's transition to independent OA publishing is inevitable. However, the study mainly provided recommendations for top-level decision-makers and had not identified the actual challenges that China's OA journals need to overcome. Zhao et al.(Zhao & Wang, 2019), Yang et al. (S. L. Yang, Xing, & Wolfram, 2018) evaluated and compared OA journals between China and USA regarding influences and impacts. Hu (Hu, 2012) evaluated availability of open access journals in the humanities and social sciences in China. On the other hand, among the studies on China's OA policy, Hu et al. (Hu, Huang, & Zhou, 2012), Shao et al.(Shao, Shen, Zhang, He, & Zheng, 2013), Ning (B. Ning, 2022) have provided detailed introductions and analyses of the current status, challenges, opportunities, and repository construction and management in the context of open access publishing in China. They all indicated that while China faces many difficulties and challenges, there are also numerous opportunities and room for development. Yang et al.(L. Yang & Ye, 2015) utilized co-word analysis and concluded that the joint efforts of government policies and academic initiatives have played a crucial role in promoting the development of open access in China. These factors have worked together to facilitate the progress of open access initiatives in the country. Therefore, it is necessary to conduct further statistical analysis on the development trends of open access journals to provide a reference for the Chinese policy makers and journal practitioners.

Regarding the different influences of OA publishing across fields, Gao (Gao, 2020) suggested that in fields such as biology, medicine, and materials science, active promotion of OA publishing should be encouraged for China, while in applied disciplines like engineering and computer science, a significant increase in the proportion of OA publications may result in the loss of high-quality papers for China. Nevertheless, these conclusions were primarily based on international development trends and did not fully consider the current state of OA development in China. According to 2023 Annual tables of the Nature Index (Woolston, 2023), China's adjusted share in the natural sciences — which includes the physical sciences, chemistry, Earth and environmental sciences and biological sciences — soared by more than 21% from 2021 to 2022, surpassing the United States for the first time in 2023, although there is still room for improvement in the biological sciences, a category formerly referred to as the life sciences in the Nature Index. Considering China's significant investment in fundamental research fields such as life sciences and medicine, we can never over-stress the necessity and importance of coordinated arrangements to enhancing China's OA journal publishing capabilities in these areas.

Therefore, only by understanding the evolving landscape of OA journal publishing can policymakers and stakeholders make informed decisions and develop strategies that contribute to the sustainable growth and improvement of open access initiatives in China. However, the research on OA journals from China has been limited to single-dimensional indicators, such as the journal impact factor (JIF) or publication capacity, which is hard to align with the needs of the Chinese scientific research. This limitation is partly due to the impractical nature of the current evaluation system for journals. When using scientometric theories to analyze journal performance and create evaluation indicators (e.g., various indicators of "highly influential international academic journals" in China), typically it is only implementable to superficially assess the excellence of journals on account of basic data (such as JIF and the number of citable items) but neglect the correlation between these data, not to mention identify specific problems faced by journals based on a solid theoretical foundation and propose practical and easily implementable technical indicators to guide journal management practices. Consequently, decision-makers may become confused, and the excessive pursuit of metrics can lead to unnecessary competition among domestic journal practitioners, overshadowing the original intention of improving publishing quality.

Based on the quantity and content of OA publications from China, we have a solid foundation to establish top-tier OA journals. However, if the designing of journal publishing models lacks innovation and conforms to prevailing practices, this can lead to internal conflicts and stagnation within the journal publishing industry. Consequently, high-quality OA articles from China may end up being published in overseas OA journals, diminishing the influence of domestic OA journals. To achieve truly valuable internationalization, OA journals from China should support their own scientists and strive to attract both domestic and international researchers.

This study can provide essential information regarding current publishing capabilities of China's international OA journals and bridge the information gap between domestic OA journal practitioners and policy makers. Ultimately, the goal should not be solely focused on acquiring editorship roles in international publishing cooperation on OA journals; instead, the emphasis should be on elevating the substantive publishing capabilities of China's OA journals themselves.

## 2. Issues with current journal evaluation system and necessity of researching elasticity

Citation impact indicators in the field of bibliometrics can be classified into two categories: size-dependent and size-independent indicators (Waltman, 2016). Given the heterogeneous nature of individual objects of study, such as diverse research fields or regions, it has become common practice to first measure the citation impact (productivity, influence scores, etc.) at the individual level and then aggregate each data to generate the citation impact at the aggregate level. Finally, the average value is calculated to eliminate the effect of size when necessary. Based on this rationale, various pairs of citation impact indicators (i.e., the size-dependent indicator and its size-independent counterpart belongs to one pair) have been developed. For example, the total number of citations received by a journal per year represents a size-dependent indicator, while the annual average number of citations per publication such as the journal impact factor (JIF) in the case of Claviate's most prominent indicator represents a size-independent counterpart.

In bibliometrics, data on publications and their received citations are typically classified as "output data." When bibliometricians devise algorithmic models to represent research productivity, the size effect of the number of publications on citations is either combined with other effects, such as article types, when calculating size-dependent indicators, or completely disregarded when calculating size-independent indicators. Consequently, journal evaluations and their relative rankings based on these indicators can be manipulated either by publishing more articles of the review type that attract additional citations or by restricting journal capacity to prevent the negative influence of rapid journal expansion on the JIF.

Furthermore, the interplay between the annual number of publications of a journal and its IF has received limited attention in the literature, despite being an important issue worth exploring. This is because publications from a journal do not yield outputs of equal value due to variations in citation counts among individual publications within the journal. In other words, if all publications within a journal were deemed of equal value, then the number of publications would be uncorrelated with the journal's IF and considered a simple output. However, this assumption is far from accurate. For instance, the correlation between the number of publications and citations highlights proliferation as a determining factor for being among the highly cited researchers (Aksnes & Aagaard, 2021). This implies that quantity and quality are inherently interconnected in research evaluation. Failing to understand this point in journal management may lead to difficulties in distinguishing between good and bad development pathways for journals; for example, a combination of either high publication yield and low JIF, or low yield and high JIF may make no differences in the outcome of evaluation of journal performances, regardless of using the size-dependent or size-independent indicators for evaluating these combinations. Decision-makers should be cautious since distorted information or an incomplete narrative provided solely by these indicators could be misleading.

On the other hand, despite the identification of several shortcomings, the H-index does not separate the number of publications from their citations when measuring scientific impact. Consequently, the H-index can provide more useful and integrative information compared to purely size-dependent or size-independent indicators. This may partially explain its popularity. However, the H-index still fails to illustrate the correlation between the number of publications and their citations. In this study, we not only considered this correlation but also went one step further to operationalize the economic concept of elasticity in the specific context of journal evaluation to strengthen the theoretical foundation of the analysis. Elasticity, a measure used in economics, focuses on the proportional effect of a change in one variable on another and is unit-free, making it suitable for comparing elasticities of journals across research fields. Economists employ elasticities to summarize various quantitative impacts of interest, and empirical works (Garcia & Thomas, 2001; Latzko, 1999) in microeconomics using elasticity measures have convincible demonstrated that valuable information can be derived, such as the development status and potential of corporations (or journals in this study), which is undoubtedly valuable for journal management.

3. Data and processing

**Data collection**

The scope of this study encompassed all international OA journals from China that are indexed in the newest Journal Citation Reports (JCR) dataset of the Web of Science (WoS) core collections (updated Oct 19, 2022). The filters were set as "Country/region: CHINA MAINLAND" and "Open Access (OA)-Article Level Filters: From 95% to 100%". It is noted that fully OA journals may not publish all items as 100% OA, and there may be other types of documents such as "subscription and free to read" and "other (non-citable items)" present, but their percentage was limited and does not significantly impact the attribute of being fully OA.

There was a total of 150 international OA journals from China indexed in the WoS core collections,

including 107 SCIE/SSCI journals with journal impact factors (JIFs) and 43 ESCI journals without JIFs. For comparison, this study also included 100 international OA journals with the highest JIFs from the WoS core collections. The filter used was "Open Access (OA)-Article Level Filters: From 95% to 100%", without restricting regions. As a result, 76 OA journals from overseas and 26 OA journals from China were identified among these 100 journals. The number of publications per year (PUB) and their corresponding JIFs from the initial year of OA until 2011, when the latest JIF applied, were retrieved from WoS and JCR, respectively, for the elasticity analysis.

The basic data of OA journals/publications from China and SCIE-indexed journals/publications from China were obtained from the InCites dataset, which was last updated on May 26, 2023. The InCites dataset includes Web of Science content indexed through Apr 30, 2023.

**Data processing**

Out of the 107 SCIE/SSCI OA journals from China and the 76 OA journals from overseas, 65 OA journals from China and 57 OA journals from overseas had their first JIFs before 2019, thereby having equal to or more than four data pairs of JIF and PUB, which made them eligible for the elasticity research. These journals were then labeled in descending order of JIFs from CN1 to CN65 and from W1 to W57, respectively. The corresponding titles and eISSN serial numbers are provided in the Supporting Information section.

To assess data distributions for normality, OriginPro 2018 (OriginLab Corporation, Northampton, MA, USA) was used. Excel computer spreadsheets (Microsoft, Seattle, WA, USA) were the primary tool for conducting calculations in this study. A P-value less than 0.05 was considered statistically significant.

**4. Construction process of economic model of elasticity**

We employed a direct approach to verify the economic model that explains the journal impact factor (JIF) using the quantifiable variable of the number of publications each year (PUB) for an OA journal. This approach is based on three fundamental elements (Rochet & Fraysse, 1997).

(1) Ceteris paribus assumption: While researchers recognize the existence of various external forces beyond PUB (e.g., retractions of publications due to scientific fraud, changes in article processing charges of publications, or shifts in researchers' focus on journal content), these factors are held constant in the model construction. It is important to emphasize that this does not imply that other factors do not influence JIFs; rather, these variables are assumed to remain unchanged during the study period.

(2) Utility maximization: We assumed that decision-makers, such as journal editors and supervisors, strive to optimize PUBs to maximize utility. This assumption forms the mathematical foundation of our economic model. Journal editors and supervisors are considered as JIF-takers. The variable outside the decision-maker's control is referred to as the exogenous variable. Since JIFs are determined independently of the editor's behaviors, and our objective is to examine how journal editors adapt to JIFs by adjusting the quantities of publications each year (i.e., studying JIF(PUB)), the JIF serves as the exogenous variable. The outcome of such decisions, such as the quantities of publications every year, represents an endogenous variable determined within our model.

The ceteris paribus assumption is enforced by manipulating only one exogenous variable while keeping all others constant. The empirical proposition for the utility-maximization hypothesis of a journal is then generated as follows:

$$\text{Utility}=\pi=\text{PUB}\times\text{JIF(PUB)}-C(\text{PUB}) \qquad \text{Eq. 1}$$

Suppose that a journal publishes all the desired documents at a certain JIF, and the total costs of publishing activity, denoted as *C*, depend on the annual number of publications (PUB). In this model, we have only one endogenous variable—PUB, which represents the quantity the journal chooses to publish. Additionally, there is only one exogenous variable—JIF, which the journal considers as given. Our model predicts how changes in this exogenous variable influence the journal's decision on PUB. It is important to note that for an OA journal, our model applies to the time period when both JIF and PUB are present after the journal becomes a full OA journal, as this ensures the satisfaction of the ceteris paribus assumption.

(3) It is important to make a careful distinction between "positive" and "normative" questions. So far, our focus has been on positive economics, specifically examining the actual correlation between PUBs and JIFs for OA journals. Our predictions are solely based on observations and do not involve personal views. However, to address real-world implications, some economists also employ normative analysis to take a definitive stance on what should be done. Particularly in the present study, a normative question regarding how PUBs should be planned for a certain journal would be based not only on the results of the model but also on the economist's personal view. But this study will only discuss the results solely based on observations.

Then, the elasticity of $e_{\text{PUB,JIF}}$ is defined as the percentage change in the number of publications per year demanded that results from a 1 percent change in JIF as shown in Eq. 2:

$$e_{\text{PUB,JIF}} = \frac{d\text{PUB}}{d\text{JIF}} \times \frac{\text{JIF}}{\text{PUB}} \qquad \text{Eq.2}$$

then we calculate the arc elasticity of the 112 target journals using Eq. 3:

$$e_{\text{PUB}n,\text{JIF}} = \frac{\text{PUB}_n - \text{PUB}_{n-1}}{\text{JIF}_n - \text{JIF}_{n-1}} \times \frac{\text{JIF}_n + \text{JIF}_{n-1}}{\text{PUB}_n + \text{PUB}_{n-1}} \qquad \text{Eq.3}$$

where the subscript *n* represents a certain year and the subscript (*n*−1) represents the year before the certain year. Then, for a data set of *n* year's data pairs of JIF and PUB, (*n*−1) of $e_{\text{PUB}n,\text{JIF}}$ would be generated for elasticity analysis.

For utility maximization, it should be satisfied according to Eq. 4 that:

$$\frac{d\pi}{d\text{PUB}} = \frac{d(\text{PUB} \times \text{JIF}(\text{PUB}))}{d\text{PUB}} - \frac{dC(\text{PUB})}{d\text{PUB}} = \text{JIF} + \text{PUB} \times \frac{d\text{JIF}}{d\text{PUB}} - \frac{dC(\text{PUB})}{d\text{PUB}} = 0 \qquad \text{Eq. 4}$$

then

$$\text{JIF} + \text{PUB} \times \frac{d\text{JIF}}{d\text{PUB}} = \frac{dC(\text{PUB})}{d\text{PUB}} \qquad \text{Eq. 5}$$

where $\frac{dC(\text{PUB})}{d\text{PUB}}$ is the marginal cost (MC), and marginal revenue in microeconomics can be translated to $\text{JIF} + \text{PUB} \times \frac{d\text{JIF}}{d\text{PUB}}$, which is the change in total JIF resulting from a change in the number of publications per year. And if MR is expressed in terms of JIF and $e_{\text{PUB,JIF}}$, it will be as Eq. 6.

$$\text{MR} = \text{JIF}\left(1 + \frac{\text{PUB}}{\text{JIF}} \times \frac{d\text{JIF}}{d\text{PUB}}\right) = \text{JIF}\left(1 + \frac{1}{e_{\text{PUB,JIF}}}\right) \qquad \text{Eq. 6}$$

As long as $e_{\text{PUB,JIF}} < 0$, marginal revenue will be less than JIF. And the circumstances with $e_{\text{PUB,JIF}} < 0$ can be further divided into two categories. (1) If demand is elastic ($e_{\text{PUB,JIF}} < -1$), then MR will be positive, which means the publishing of one more document will not affect JIF remarkably, and hence more revenue will be yielded by publishing more documents. It should be noted that if demand facing the journal is infinitely elastic ($e_{\text{PUB,JIF}} = -\infty$), MR will equal JIF, in which case the journal is a JIF-taker and PUB has no influence on JIF. (2) Conversely, if demand is inelastic ($e_{PUB,JIF} > -1$), MR will be negative. Increases in PUB can only be achieved by substantial decreases in JIF, and these decreases will result in decreases in total utility (i.e., the product of PUB and JIF).

We can also rearrange Eq. 5 to determine the JIF−MC markup by using Eq. 7.

$$\frac{JIF-MC}{JIF} = \frac{1}{-e_{PUB,JIF}} \qquad \text{Eq. 7}$$

Eq. 7 implies that the percentage markup over MC will be higher as $e_{PUB,JIF}$ is closer to −1. For a perfect competition market, where there are many other journals publishing documents of analogous quality, then $e_{PUB,JIF} = -\infty$, indicating no makeup (JIF=MC). However, with an $e_{PUB,JIF}=-5$, the markup over MC will be 20% of JIF since (JIF−MC)/JIF=0.2.

Assuming the demand function linear over the range of interest, then the optimized JIF ($JIF_{opt}$) and PUB ($PUB_{opt}$) can also be obtained by using the demand function based on $\bar{e}_{PUB,JIF}$, $\overline{PUB}$ and $\overline{JIF}$ as shown in Eq. 8:

$$PUB = \overline{PUB}[1+\bar{e}_{PUB,JIF}(JIF-\overline{JIF})/\overline{JIF}] \qquad \text{Eq. 8}$$

where $\bar{e}_{JIF,PUB}$ represents the average of $(n-1)$ $e_{PUBn,JIF}$ values that have been calculated from Eq. 3, and $\overline{PUB}$ and $\overline{JIF}$ denote the average PUB and JIF, respectively, within the time scope of our model. Specifically, this refers to the period when $JIF_n$ and $PUB_n$ are both present after the journal has transitioned to a full open access (OA) journal.

Transform Eq. 8 to obtain JIF as a function of PUB as shown in Eq. 9. According to Eq. 9, the demand curve will be negatively sloped if $\bar{e}_{PUB,JIF}<0$. Also derive MR as a function of PUB as shown in Eq. 10, and it is shown in Fig. 1 that the MR curve would fall below the demand curve.

$$JIF = \frac{\overline{JIF}}{\overline{PUB} \times \bar{e}_{PUB,JIF}} \times PUB + \overline{JIF}\left(1 - \frac{1}{\bar{e}_{PUB,JIF}}\right) \qquad \text{Eq. 9}$$

$$MR = \frac{2\overline{JIF}}{\overline{PUB} \times \bar{e}_{PUB,JIF}} \times PUB + \overline{JIF}\left(1 - \frac{1}{\bar{e}_{PUB,JIF}}\right) \qquad \text{Eq. 10}$$

Using journal No. W55 as an example, the demand curve and MR curve can be shown in Fig. 1. For annual numbers of publications larger than $PUB_{opt}$, the MR is negative. At $PUB_{opt}$, the total revenue (or utility) $PUB_{opt} \times JIF_{opt}$ is maximized. Beyond this point, further increases in PUB would lead to a decrease in total utility, as the positive impact of PUB on total utility becomes outweighed by the simultaneous decrease in JIF.

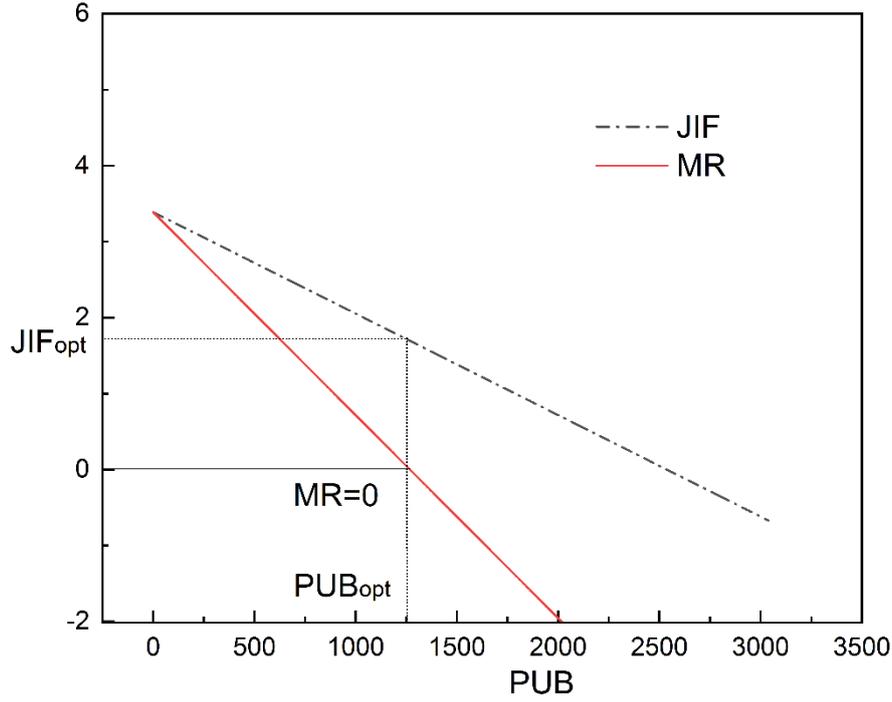

**Figure 1** Demand curve and associated marginal revenue curve for journal No. W55, where $\bar{e}_{\text{PUB,JIF}} = -28.621$, $\overline{\text{PUB}} = 85.636$, and $\overline{\text{JIF}} = 3.274$.

Then for the cases where $\bar{e}_{\text{JIF,PUB}} < 0$ and the utility maximization hypothesis applies, we will estimate JIF$_{\text{opt}}$ and PUB$_{\text{opt}}$ at MR=0 and $\bar{e}_{\text{JIF,PUB}} = -1$ by using Eq. 11 and Eq. 12.

$$\text{JIF}_{\text{opt}} = \left(\frac{1}{2} - \frac{1}{2\bar{e}_{\text{PUB,JIF}}}\right) \times \overline{\text{IF}} \qquad \text{Eq. 11}$$

$$\text{PUB}_{\text{opt}} = \frac{1 - \bar{e}_{\text{PUB,JIF}}}{2} \times \overline{\text{PUB}} \qquad \text{Eq. 12}$$

## 5. Results

**Current status of OA journal publishing in China**

Data from the InCites Dataset as shown in Table 1 reveal a consistent increase in the share of OA publications from China in total SCIE contents from China, rising from 30.6% in 2017 to 44.9% in 2022. This represents a significant increment of 47% for OA publications from China. Given the widely acknowledged higher citation impact of OA publications compared to traditional non-OA publications, China should proactively explore the immense potential of its OA scholarly publishing needs. However, a critical concern remains regarding the insufficient number of publishing channels for OA publications from China, compared to the substantial demand for such publications driven by China's research capacity. The ratio of SCI$_{\text{OA-J}}$/SCI$_{\text{OA}}$, which stands at 6% and has not seen a boost from 2017 to 2021, highlights this gap. Despite the doubling of OA publications published in academic journals belonging to China from 8537 in 2017 to 19924 in 2022 and a consistent increase in the number of OA journals from China aligning with the global OA journal industry, these developments do not represent a substantial enhancement in the OA publication capacity of academic journals in China relative to China's ever-growing academic strength. Consequently, an outflow of OA publications overseas is inevitable within the context of a more mature global OA publishing framework.

Table 1 Comparisons of basic data information on a yearly basis between OA journals/publications of China and SCIE-indexed journals/publications of China from 2017 to 2022. Data retrieved from Incites Dataset on June 10, 2023

| Year | Number of SCI OA publications from China ($SCI_{OA}$) | Number of SCI publications from China ($SCI_{China}$) | $SCI_{OA}/SCI_{China}$ | Number of OA publications in China's journals ($SCI_{OA-J}$) | Number of publications in China's journals ($SCI_J$) | $SCI_{OA-J}/SCI_{OA}$ | $SCI_{OA-J}/SCI_{China}$ |
|---|---|---|---|---|---|---|---|
| 2017 | 141389 | 461602 | 30.6% | 8537 | 24001 | 6.0% | 35.6% |
| 2018 | 161901 | 506764 | 31.9% | 9012 | 25000 | 5.6% | 36.0% |
| 2019 | 205875 | 597130 | 34.5% | 11548 | 27095 | 5.6% | 42.6% |
| 2020 | 245124 | 633153 | 38.7% | 17078 | 34044 | 7.0% | 50.2% |
| 2021 | 284063 | 711085 | 39.9% | 20631 | 39755 | 7.3% | 51.9% |
| 2022 | 347045 | 773206 | 44.9% | 19924 | 40613 | 5.7% | 49.1% |

Therefore, it is imperative for academic journals in China to urgently enhance their publishing capacity. The scalability of this improvement depends on two key factors: the number of OA journals and the average number of publications per journal. Currently, several challenges are associated with these two factors. Firstly, it could be beneficial if each publishing unit in China leverages its resources to launch new journals. Presently, according to Ren et al.(REN et al., 2023), the supervising units, sponsoring units, and publishing units of the 5071 academic journals in China are widely dispersed, resulting in resource redundancy and internal competition. There are 1325 supervising units, averaging 3.83 journals per unit; 3168 primary sponsoring units, averaging 1.60 journals per unit; and 4354 publishing units, averaging 1.16 journals per unit. However, as will be discussed in this study, the dispersion of supervising and sponsoring units does not necessarily imply that expanding the number of journals should be discouraged. Secondly, although the number of academic journals in China has increased, the average number of published articles per journal had been declining until 2019, only experiencing an increase in 2020 and 2021. Thus Ren et al. (REN et al., 2023) suggested that further optimizing the academic positioning and disciplinary landscape of English-language journals at the level of journals and publishing units is essential to meet the needs of academic development and cross-field exchange. However, detailed suggestions for each journal or publishing unit are not readily available in literature.

The reason for these notable discrepancies is that the majority of English-language journals from China adopt a collaborative model with overseas publishers, utilizing their publishing platforms for global dissemination. The publication scale is somewhat constrained by these overseas collaborators, particularly for journals employing the "Open Access" operational model (Zhang, Wei, Huang, & Sivertsen, 2022), where the number of published articles directly influences the fees paid by the editorial office to overseas partners (Bi Ning et al., 2021). However, there is a dearth of detailed studies specifically on the number of publications per OA journal in the literature, except for a few studies on English-language journals from China as mentioned above. Such analysis is essential for comprehending the current status of OA journals from China and for laying the groundwork for their future development.

In summary, despite the constant establishment of new journals, China's publishing capacity has not significantly increased due to a lack of national-level planning. To better understand the challenges facing OA journals in China, it is crucial to analyze the number of journals and the average number of publications per journal, respectively. Therefore, this study conducted detailed statistical analyses of OA journals from China to provide decision-makers with valuable insights for future strategies in managing these journals.

**Comparisons of OA journals from China and overseas in terms of elasticity**

To calculate the elasticity of the journal impact factor (JIF) with respect to the number of publications each year (PUB) for an OA journal (denoted as $e_{PUB,JIF}$), we collected the JIF and PUB for each year starting from the year the journal became an open access journal as indicated in the JCR, up to 2011, which represents the year witnessing the latest available JIF. In addition, we included the top 100 international OA journals worldwide according to their JIF rankings, from highest to lowest, for comparison. Among these 100 OA journals, 26 are from China, constituting 17% of all international journals from China. This indicates the high quality of OA journals from China compared to the global average.

Prior to calculating $e_{PUB,JIF}$, we conducted a correlation analysis between JIFs and PUBs for all international OA journals from China and overseas that had at least four years of JIF and PUB data. This included 65 international OA journals from China and 57 OA journals overseas. The correlation between these two factors serves as the foundation for discussions on the elasticity of one factor with respect to the other. It is worth noting that in the normality test, some data pairs of JIFs and PUBs from the 112 OA journals did not follow a normal distribution, justifying the use of Spearman correlation coefficients in the analysis. Among the 65 Chinese OA journals and the 57 OA journals overseas, 33 Chinese OA journals and 39 OA journals overseas displayed a significant correlation ($P < 0.05$) between JIFs and PUBs. This suggests that the JIFs of OA journals from China exhibit more distinct distribution behaviors in relation to their PUBs compared to OA journals overseas.

The violin plots in Figure 2 depict the averaged elasticity ($\bar{e}_{PUB,JIF}$) of the journal of interest. These plots combine characteristics of a box plot and a kernel density plot, allowing for a comparison of the distribution and summary statistics of $\bar{e}_{PUB,JIF}$ between international OA journals from China and from overseas. It should be noted that both data sets do not follow a normal distribution, as confirmed by conducting the Shapiro-Wilk normality test (suitable for data with a size below 50). Consequently, the median and 1.5 interquartile range (1.5IQR) were used as indicators in summary statistics, as shown in Figure 2.

The $\bar{e}_{PUB,JIF}$ of overseas OA journals exhibits a wider range, with an $\bar{e}_{max}$ of 19.421 and an $\bar{e}_{min}$ of −28.621, compared to OA journals from China, which have an $\bar{e}_{max}$ of 11.882 and an $\bar{e}_{min}$ of −4.054. Additionally, the 1.5IQR for $\bar{e}_{PUB,JIF}$ is also wider for overseas OA journals than for those from China. It is worth mentioning that 18 out of the 65 OA journals from China (27.7%) have negative $\bar{e}_{PUB,JIF}$, whereas 23 out of the 57 OA journals from overseas (40.4%) exhibit negative $\bar{e}_{PUB,JIF}$, which fit into our economic model. A higher percentage for OA journals from overseas suggests that high-quality OA journals from overseas tend to be more market-oriented compared to

OA journals from China. On the other hand, those journals with positive $\bar{e}_{PUB,JIF}$ should be analyzed of returns to scale.

Furthermore, it is observed that OA journals from China often experience longer waiting periods before receiving their first JIF, unlike their overseas counterparts. For instance, *Chinese Journal of Aeronautics* became an OA journal in 2002 and was indexed in WoS in 1999, but it did not receive its first JIF until several years later in 2009, which was a common occurrence for OA journals from China. In contrast, overseas OA journals typically obtained their first JIF in the second year after being indexed in WoS.

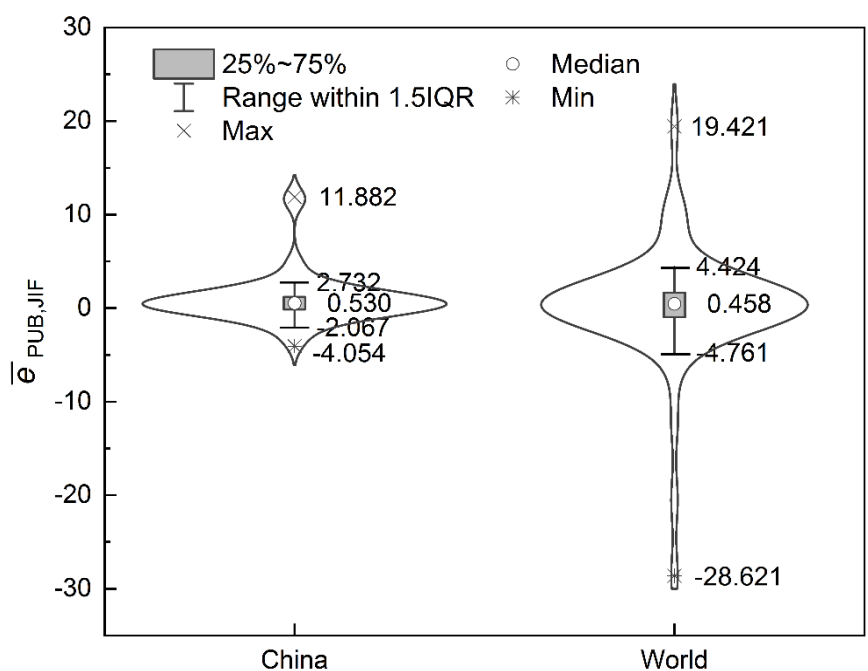

**Figure 2** Descriptive statistics of averaged elasticity ($\bar{e}_{PUB,JIF}$) of the journal impact factor (JIF) with respect to the number of publications per year (PUB) for the target journal after its open access.

**(1) Comparisons between OA journals in market setting**

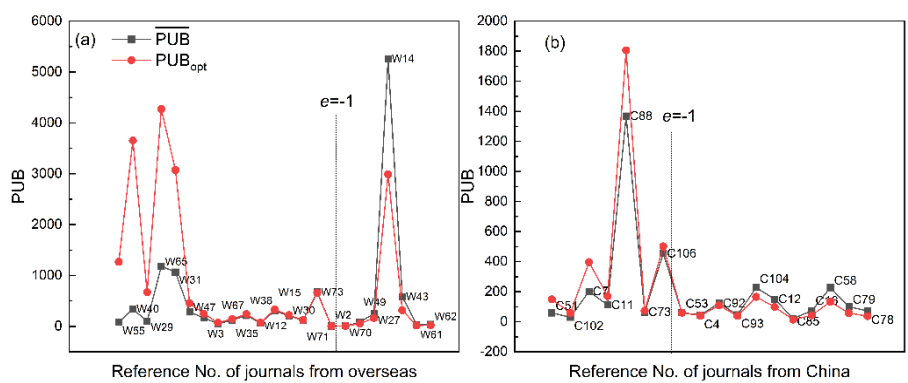

**Figure 3** Average annual number of publications ($\overline{\text{PUB}}$) and the corresponding $\text{PUB}_{\text{opt}}$ for (a) the 23 OA journals from overseas with $\bar{e}_{\text{PUB,JIF}} < 0$ and (b) the 18 OA journals from China with $\bar{e}_{\text{PUB,JIF}} < 0$. The data are sorted in ascending order based on $\bar{e}_{\text{PUB,JIF}}$ and divided into two categories using $\bar{e}_{\text{PUB,JIF}} = -1$ as the threshold.

According to Eq. 6, when $\bar{e}_{\text{PUB,JIF}} < 0$, the resulting MR will be less than JIF, as illustrated in Fig. 1. Therefore, journals with $\bar{e}_{\text{PUB,JIF}} < 0$ are of particular interest to utility-maximizing journal managers who need to understand how increases in PUB will affect JIF before making optimal PUB decisions. By computing $\text{PUB}_{\text{opt}}$ using Eq. 12, it is displayed in Fig. 3 that those journals with $\bar{e}_{\text{PUB,JIF}} < -1$ are advised to increase PUB, while those with $0 > \bar{e}_{\text{PUB,JIF}} > -1$ would benefit from reducing PUB.

The combined sum of $\overline{\text{PUB}}$ for the 18 OA journals from China from the beginning of OA to 2021 is 3436.5, compared to 11256.9 for the 23 OA journals from overseas, apparently indicating that there is still potential for OA journals from China to increase their OA publications. However, this conclusion holds true only when we disregarded the potential for reaching maximized utility, which can vary among different journals. Since $\bar{e}_{\text{PUB,JIF}} = -1$ represents the optimal point for all journals with $\bar{e}_{\text{PUB,JIF}} < 0$, the $\text{PUB}_{\text{opt}}$ of journals with $\bar{e}_{\text{PUB,JIF}} < -1$ should be larger than the corresponding $\overline{\text{PUB}}$, while the $\text{PUB}_{\text{opt}}$ of journals with $\bar{e}_{\text{PUB,JIF}} > -1$ should be smaller than $\overline{\text{PUB}}$, as shown in Fig. 3. Out of the 23 OA journals from overseas, sixteen have $\bar{e}_{\text{PUB,JIF}} < -1$, compared to seven out of the 18 OA journals from China, indicating that more OA journals from overseas have the potential for further expanding their publication capacities compared to their Chinese counterparts.

Using Eq. 11 and 12, $\text{JIF}_{\text{opt}}$ and $\text{PUB}_{\text{opt}}$ were obtained for all the OA journals with $\bar{e}_{\text{PUB,JIF}} < 0$. To compare the overall potential for development between the 23 OA journals from overseas and the 18 OA journals from China with respect to $\text{PUB}_{\text{opt}}$ and $\text{JIF}_{\text{opt}}$, the combined sums of $\overline{\text{PUB}}$ and $\text{PUB}_{2021}$ were also calculated for both groups. These values, expressed as a proportion of $\text{PUB}_{\text{opt}}$, are listed in Table 2. The results reveal that although both groups have $\text{PUB}_{\text{opt}}$ larger than $\overline{\text{PUB}}$, the gap between $\text{PUB}_{\text{opt}}$ and $\overline{\text{PUB}}$ is greater for the OA journals from overseas at 41.0%, compared to 13.0% in China's case. This indicates that OA journals from overseas with negative $\bar{e}_{\text{PUB,JIF}}$ have a greater potential for development than their counterparts in China. It is particularly noteworthy that the combined sum of $\text{PUB}_{2021}$ for the 18 OA journals from China already exceeds that of $\text{PUB}_{\text{opt}}$ by 12.2%, suggesting that further increases in PUB from 2021 on would come at the expense of substantial decreases in JIF. This conclusion is further supported by the fact that the combined sum of $\text{JIF}_{2021}$ as a proportion of that of $\text{JIF}_{\text{opt}}$ for OA journals from China is only 69.9%, significantly lower than the value for OA journals from overseas (100.1%).

In conclusion, OA journals from overseas demonstrate stronger growth momentum in the global market compared to those from China both in journal and publication quantities. Therefore, these market-oriented OA journals from China should prioritize improving the quality of their OA publications rather than focusing solely on increasing quantity, since the combined sum of PUB in 2021 in fact has already passed the optimum.

**Table 2** Comparisons of 2021 and averaged PUB, and 2021 and averaged JIF with respect to optimized PUB and JIF, between OA journals from China and overseas with $\bar{e}_{\text{PUB,JIF}} < 0$

| | Journal quantity | Sum of $\overline{\text{PUB}}$ (as a | Sum of $\text{PUB}_{2021}$ | Sum of $\overline{\text{JIF}}$ (as a | Sum of $\text{JIF}_{2021}$ (as |

|  | with $\bar{e}_{\text{PUB,JIF}} < -1$( as a proportion of journals with negative $\bar{e}_{\text{PUB,JIF}}$) | proportion of the sum of $\text{PUB}_{\text{opt}}$) | (as a proportion of the sum of $\text{PUB}_{\text{opt}}$) | proportion of the sum of $\text{JIF}_{\text{opt}}$) | a proportion of the sum of $\text{JIF}_{\text{opt}}$) |
|---|---|---|---|---|---|
| OA journals from China | 7(38.9%) | 3436.4(87.0%) | 4435(112.2%) | 193.8(50.9%) | 135.4(69.9%) |
| OA journals from overseas | 16(69.6%) | 11256.9(59.0%) | 15999(83.8%) | 377.1(51.3%) | 337.5(100.1%) |

**(2) Comparisons between OA journals by using returns to scale**

For journals with positive $\bar{e}_{\text{PUB,JIF}}$, indicating that PUB and JIF change in the same direction, the question of returns to scale arises, examining how the output (JIF) responds to increases in all inputs (in this case, PUB) together. Conceptually, two forces come into play when all inputs are doubled (Rochet & Fraysse, 1997). The first is greater division of labor and specialization of function. This may lead to increased efficiency and a JIF that is more than doubled, indicating increasing returns to scale ($\bar{e}_{\text{PUB,JIF}} > 1$). Achieving this requires implementing rigorous peer review systems, improving editorial and production standards, and other measures aimed at enhancing the service quality of journals.

The second is more complicated managerial oversight. Expanding the publication capacity of a journal can introduce complexities that may result in declining efficiency and a JIF that is less than doubled, indicating decreasing returns to scale ($0 < \bar{e}_{\text{PUB,JIF}} < 1$). Managing this situation requires addressing challenges associated with the increased workload and ensuring effective oversight.

Empirically investigating which of these tendencies becomes dominant can shed light on the dynamics of journals with positive $\bar{e}_{\text{PUB,JIF}}$. By examining the relationship between PUB and JIF for such journals, it is possible to assess whether the returns to scale lean towards increasing or decreasing returns.

Among the 47 OA journals from China depicted in Fig. 4, 28 of them, or 59.6%, have $0<\bar{e}_{\text{PUB,JIF}}<1$. In comparison, among the OA journals from overseas, 17 out of the 34 journals, or 50.0%, fall into this category. Although the difference in terms of journal quantities may not be significant, it becomes more apparent when considering publication capacities. The combined sum of averaged annual publications $\overline{\text{PUB}}$ for the 28 OA journals from China with $0<\bar{e}_{\text{PUB,JIF}}<1$ is 5330.8, accounting for 67.1% of all OA publications from China's OA journals with $\bar{e}_{\text{PUB,JIF}}>0$. On the other hand, the combined sum of $\overline{\text{PUB}}$ for the 17 OA journals from overseas with $0<\bar{e}_{\text{PUB,JIF}}<1$ is 4878.2, representing 47.2% of all overseas OA publications from journals with $\bar{e}_{\text{PUB,JIF}}>0$.

Therefore, concerning journals with positive $\bar{e}_{\text{PUB,JIF}}$, it is evident that the impact of diminishing returns to scale is more pronounced for OA journals from China compared to those from overseas. This implies that the 28 OA journals with $0<\bar{e}_{\text{PUB,JIF}}<1$ could benefit significantly from operating economies of scale, suggesting that merging certain journals could be a viable option. Merging these journals can reduce opportunity costs and marginal costs resulting from information asymmetry when journals are managed separately, while also leveraging the technological advancements offered by integrated journal management systems.

However, it is important to note that not all journals would derive substantial benefits from merging, especially for those journals with $\bar{e}_{PUB,JIF}>1$ that exhibit significant increasing returns to scale. In such cases, seeking efficiencies through expanding the publication capacity of individual journals would be more desirable. This expansion would enable the journal management to align with more refined divisions of specialization, taking advantage of the opportunities presented by increased scale.

In conclusion, this analysis can provide insights into the variable impacts of expanding the publication capacity of different journals on JIF and guide decision-makers through the management and growth strategies of these journals on a case-by-case basis; e.g., merging journals concerning the same cutting-edge topics or belonging to the same publishing units can be proposed to meet the needs of academic development and cross-field exchange.

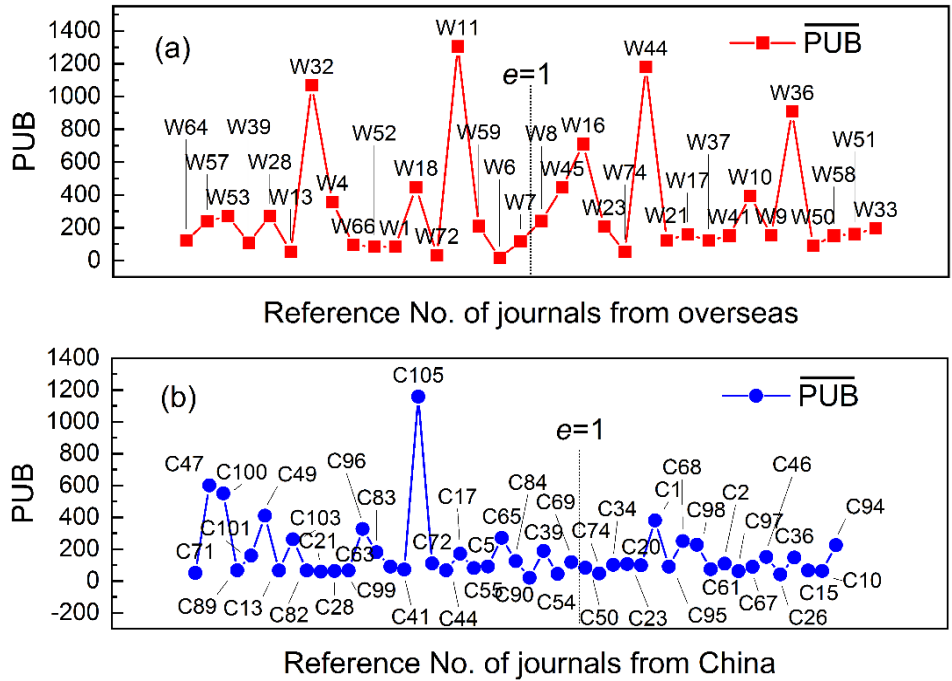

**Figure 4** Average annual number of publications ($\overline{PUB}$) from the beginning of OA to 2021 for (a) the 34 OA journals from overseas with $\bar{e}_{PUB,JIF} > 0$ and (b) the 47 OA journals from China with $\bar{e}_{PUB,JIF} > 0$. The data are sorted in ascending order based on $\bar{e}_{PUB,JIF}$ and divided into two categories using $\bar{e}_{PUB,JIF}=1$ as the threshold.

## 6. Discussion

**Table 3** The number of OA journals and the averaged $\overline{PUB}$ per journal (the combined sum of $\overline{PUB}$/ journal quantities) classified according to the values of $\bar{e}_{PUB,JIF}$ for China and overseas, respectively

|  | Journal quantities from | Combined sum of $\overline{PUB}$/ journal | Journal quantities from | Combined sum of $\overline{PUB}$/ journal |
| --- | --- | --- | --- | --- |

|  | China | quantities from China | overseas | quantities from overseas |
| --- | --- | --- | --- | --- |
| $\bar{e}_{\text{PUB,JIF}} \leq -1$ | 7 | 327.1 | 14 | 308.0 |
| $-1 < \bar{e}_{\text{PUB,JIF}} \leq 0$ | 11 | 104.3 | 9 | 711.6 |
| $0 < \bar{e}_{\text{PUB,JIF}} < 1$ | 28 | 197.5 | 17 | 286.9 |
| $\bar{e}_{\text{PUB,JIF}} \geq 1$ | 19 | 126.7 | 17 | 320.5 |
| Sum | 65 | - | 57 | - |

In 2021, the average annual number of publications per journal for all SCIE-indexed journals from China was 151, with a median of 109 (REN et al., 2022). In comparison, that of the 65 OA journals from China in this study was 250.2, with a median of 140. While OA journals from China exhibit a higher average annual number of publications per journal compared to all SCIE-indexed journals, this does not necessarily imply that the publication capacities of OA journals from China are abundant to meet the academic demands of domestic researchers for OA publishing activity, which is also displayed in Table 1, particularly considering the gap in annual OA publication capacities between China and overseas. Furthermore, OA journals typically publish a larger quantity of papers than traditional journals, indicating the desirability of further increasing publication capacities for OA journals from China. Additionally, since the combined sum of OA publications depends not only on the number of OA journals but also on the annual average number of publications per journal, increasing the number of OA journals without establishing new supervising units, primary sponsoring units, or publishing units can contribute to the overall increase in OA publications.

It is essential to carefully consider the future trajectory of OA journals from China, taking into account the elasticity of PUB with respect to JIF, which is closely tied to the development stage of each individual journal.

(1) For the 7 journals from China with $\bar{e}_{\text{PUB,JIF}} \leq -1$, they demonstrate great momentum for growth in the context of global publishing market, but the averaged $\overline{\text{PUB}}$ per journal already surpassed that of 14 journals from overseas. Therefore, it is viable to launch new OA journals that are guided by market forces within this category; however, an excessive pursuit of further increasing publication capacity of the existing 7 journals (either due to government intervention or the personal choices of journal managers) is not recommended. Journal managers should not be apprehensive about the natural decline in JIF resulting from increased PUB for journals belonging to this category, as this signifies an active journal with growth potential in the market setting.

(2) In the range of $-1 < \bar{e}_{\text{PUB,JIF}} \leq 0$, OA journals from China have the smallest averaged $\overline{\text{PUB}}$ per journal among 4 categories. On the other hand, overseas OA journals have a largest averaged $\overline{\text{PUB}}$ per journal of up to 711.6 in the same category. As indicated by the elasticity analysis, overseas OA journals should undoubtedly reduce the annual numbers of publications, while it is more feasible for OA journals from China to consider merging two or more journals within this category due to their relatively small averaged $\overline{\text{PUB}}$ per journal, which should not be further decreased, although decreasing in publication capacity is implied in the elasticity analysis.

(3) With $0 < \bar{e}_{\text{PUB,JIF}} < 1$, OA journals from China exhibit significant diminishing returns to scale and have a larger number of journals compared to other categories. It implies that the overall effect of increasing the annual publication capacity on these journals is negative, such as problems arising from coordinating and supporting a larger number of academic editors, editorial board members, and reviewers from different disciplines, and balancing the inconsistency in the criteria applied by different academic editors and reviewers. Therefore, it is reasonable to merge journals A and B in order to minimize marginal costs while possibly keeping the total cost unchanged, i.e., maximizing $PUB_A \times JIF_A + PUB_B \times JIF_B$, subject to a constant budget constraint of $C(A,B)=C$. In practice, the benefits of merging are exhibited by NPG's *Nature Communications*, *Communications Physics*, *Scientific Reports*, and other similar high-ranking OA journals that adopt a cascading peer review model, which developed robust editorial processes and ensured efficient publication workflows.

(4) For $\bar{e}_{PUB,JIF} \geq 1$, it should be noted that $|\bar{e}_{PUB,JIF}| \geq 1$ (i.e., both $\bar{e}_{PUB,JIF} \leq -1$ and $\bar{e}_{PUB,JIF} \geq 1$) is the circumstances that are favorable for journal management, since JIF can be substantially influenced by PUB. However, OA journals from China have not fully utilized this influence to expand the publication capacities of journals with $\bar{e}_{PUB,JIF} \geq 1$, as evidenced by an averaged $\overline{PUB}$ per journal of only 126.7, compared to 320.5 for OA journals from overseas. In fact, OA journals in this category are in a better position than their counterparts with $\bar{e}_{PUB,JIF} \leq -1$, as JIF increases, rather than decreases, alongside PUB, which is potentially attributed to various policy supports that incentivize the growth of publication capacity of a certain journal. Therefore, OA journals in this category should prioritize steadfast expansion of publication capacity over consistent growth of JIF.

In summary, based on the development and evolution of journals such as *PLOS ONE*, *Scientific Reports*, and *IEEE Access*, Petrou(Petrou, 2020) categorized the lifecycle of Open Access Mega Journals (OAMJs) into four stages: accumulation, rapid growth, turbulence, and slow decline. Due to the dynamic nature of journals throughout their lifecycle, determining the role and future trajectory of an individual OA journal in academic publishing is a complex task influenced by various factors, including governmental and funding agency support. Consequently, there is limited research available to offer decision-making recommendations for journal practitioners. This study pioneers an exploration of the underlying mechanisms that govern the relationship between annual publication capacity (PUB) and output quality (JIF), providing insights into enhancing the conversion efficiency from PUB to JIF for individual journals through the lens of microeconomic theory. The findings suggest that leveraging the distinctive attributes of OA journals (such as their substantial publication capacity and swift publication speed), supported by appropriate management models and a primary focus on the "scientific soundness" principle in manuscript evaluation rather than the "novelty" or "importance" principles emphasized by traditional journals(REN, Gao, & Cheng, 2020), can significantly improve the conversion efficiency from PUB to JIF ($|\bar{e}_{PUB,JIF}| \geq 1$). Conversely, if the advantages of large publication capacity and fast publication speed are not effectively harnessed, and if the expansion of PUB is not accompanied by improvements in management models, but resulting in a decline in manuscript evaluation standards, the efficiency of transitioning PUB to JIF ($0<|\bar{e}_{PUB,JIF}|<1$) may be compromised. This study sheds light on the distinct role and position of individual OA journals in academic publishing, offering valuable guidance for decision-makers and indicating the development directions that OA journals from China should pursue to contribute more effectively to scientific advancement.

7. Conclusion

The interplay between annual publication capacity and impact factor (IF) has received limited attention in the literature, despite its significance in research evaluation. To facilitate strategic decision-making in the development of OA journals from China, it is crucial to begin by identifying the specific challenges encountered by these journals through rigorous theoretical analysis; therefore, this study aimed to assess the performance of open access journals from China using microeconomic analysis, focusing on the elasticity of annual publication capacity with respect to the journal impact factor. This analysis provides essential insights into the current publishing capabilities of China's OA journals and bridges the information gap between domestic OA journal practitioners and policy makers.

Subsequently, practical and easily implementable technical indicators should be proposed, which can effectively guide journal management practices. For OA journals with $\bar{e}_{\text{PUB,JIF}} \leqslant 0$ operating within a market framework, the JIF-MC markup can be generated for each journal by using Eq. 7 and the optimal JIF and annual publication capacity can be computed by using Eq. 11 and Eq. 12, and the results are exhibited in Fig. 3. Moreover, overseas journals are demonstrated to have a greater potential for increasing their annual publication capacities without significant declines in their impact factors. Conversely, OA journals from China witness that their annual publication capacities are nearing saturation. Further increases in annual publication quantity (PUB) may lead to substantial decreases in the impact factor (JIF).

Furthermore, OA journals with $0 < \bar{e}_{\text{PUB,JIF}} < 1$ display diminishing returns to scale, indicating that merging specific journals within this category can optimize costs and enhance overall efficiency. On the other hand, OA journals with $\bar{e}_{\text{PUB,JIF}} \geqslant 1$ present an opportunity for expanding publication capacities; thus it is crucial for China to leverage this potential and focus on increasing annual publication capacity of these journals to maintain competitiveness with overseas counterparts.

These findings in the present study highlight the importance of considering the elasticity of annual publication capacity with respect to the impact factor in journal evaluation. While it is widely-acknowledged that upholding publication quality remains paramount, striking a balance between quantity and quality is also urgently needed, and this study provides a reference for individual journals and publishing units about how to better meet the academic demand from domestic researchers. Ultimately, the goal is to foster the growth and success of OA journals in China by addressing specific challenges and adopting evidence-based approaches. By doing so, the academic publishing units in China can enhance its contributions to the broader scholarly ecosystem and promote the dissemination of high-quality research.